# Using Multi-Source Data to Identify High-Emitting Heavy-Duty Diesel Vehicles

Zhuoqian Yang[a], Ke Han[b,*], Linwei Liao[a], Jiaxin Wu[a]

[a]School of Transportation and Logistics, Southwest Jiaotong University, China

[b]School of Economics and Management, Southwest Jiaotong University, China

[*]Corresponding author. Email: kehan@swjtu.edu.cn

## ABSTRACT

Identifying and managing high-emitters among heavy-duty diesel vehicles is a key to mitigating urban air pollution, as a small number of such vehicles could contribute a significant amount of total transport emissions. On-board monitoring (OBM) systems can directly monitor the real-time emission performance of heavy-duty vehicles on road and have become part of the future emissions compliance framework. The challenge, however, lies in the frequent unavailability of OBM data, affecting the effective screening of high-emitting vehicles. This work proposes to bridge the gap by integrating OBM data with remote sensing data to create a comprehensive monitoring system. OBM data is used to characterize the detailed real-world $NO_x$ emission performance of both normally-behaving vehicles and high-emitters at various vehicle operating conditions. Remote sensing data is employed to screen out candidate high-emitting vehicles based on thresholds determined by OBM data. Finally, the dynamic $NO_x$ emission reduction potential across all roads is mapped by combining the trajectory data for each vehicle with the emission data. A case study in Chengdu, China, utilizing emission and traffic data from heavy-duty vehicles for transporting construction waste (a.k.a. slag trucks), reveals the national threshold for identifying high-emitters via remote sensing might be too lenient, particularly in the medium speed range. An emission reduction of 18.8% in the China V slag truck fleet could be achieved by implementing this novel method in practice in Chengdu. This approach establishes a reliable and ongoing scheme for pinpointing high-emitters through multi-source data, which allows local authorities to develop more robust and targeted strategies to mitigate urban air pollution from heavy-duty diesel vehicles.

# 1. Introduction

Road transport is a major emission source of Nitrogen Oxides ($NO_x$) (Pastorello and Melios, 2016). Specifically, vehicles on road contribute 58.9% of the total $NO_x$ emissions in China (MEE, 2022). Among all the vehicle types, heavy-duty vehicles are responsible for 76.1% of the $NO_x$ in the road traffic sector (MEE, 2022), despite comprising only 2.9% of vehicles on road (NBS, 2023). A predominant 96.4% of heavy-duty vehicles (MEE, 2019) are powered by diesel, and researchers (Bishop et al., 2016, Huang et al., 2018, Yang et al., 2022) have found that a small number of high-emitting diesel vehicles with manipulated and defective $NO_x$ emission reduction systems could contribute a significant amount of total $NO_x$ emissions. Accordingly, identifying high-emitters in heavy-duty diesel vehicles (HDDVs) is a key to mitigate $NO_x$ pollution.

Diverse methods have been developed to monitor the real-world emissions from vehicles and screen out high-emitters (Ropkins et al., 2009, Franco et al., 2013). The main emission measurement techniques include laboratory (chassis dynamometer) tests (Demuynck et al., 2012, Moody and Tate, 2017), on-board tests utilizing portable emissions measurement systems (PEMS) (O'Driscoll et al., 2016, Luján et al., 2018), remote sensing technology for vehicle emission testing (Carslaw et al., 2011, Chen and Borken-Kleefeld, 2016), and on-board monitoring (OBM) systems (Zhang et al., 2020, Müller et al., 2022). Laboratory tests and PEMS provide second by second emission rates (grams.sec$^{-1}$) over a whole driving cycle, ideal for vehicle type approvals but are time-consuming and expensive even for small samples of vehicles. Remote sensing devices (RSDs), however, can un-intrusively take a snap-shot sample of fuel-specific emission rates (grams.kg$^{-1}$) from a large number of vehicles in a single day (Beaton et al., 1995, Huang et al., 2018), proving cost-effective for fleet emission monitoring (Carslaw et al., 2011, Carslaw et al., 2013, Chen and Borken-Kleefeld, 2016, Grange et al., 2019) and high-emitter detection (Borken-Kleefeld, 2013, Pujadas et al., 2017, Huang et al., 2019). Nonetheless, some research contends that the validity of a single RSD record in reflecting the actual emission performance of a passing vehicle requires assessment (Qiu and Borken-Kleefeld, 2022). OBM systems, which simultaneously monitor $NO_x$ emissions and gather vehicle performance data from preceding on-board diagnostics (OBD) systems, have recently been incorporated into the emissions compliance framework (Barbier et al., 2024). OBM-based calculation models have been developed to assess the $NO_x$ emission performance of HDVs (Yang et al., 2016, Zhang et al., 2020). Yet, due to sensor malfunctions or failures in the data transmission process, the emission-related data from OBM for many HDDVs is unavailable (Li et al., 2023) or missing (Barbier et al., 2024), preventing effective monitoring of high-emitting vehicles. In conclusion, while numerous technologies are available for monitoring vehicle emission levels and identifying high-emission vehicles, concerns persist regarding cost, data accessibility and reliability.

Since relying on singular emission data falls short of meeting the needs for precise regulation, researchers have emphasized the importance of building a comprehensive diagnostic framework for continuous, reliable and effective emission monitoring (Hao et al., 2022). Fernandes et al. (2019) have combined PEMS data with internally observable variables from OBD to integrate second-by-second vehicle activity and emission rate predictions. Xie et al. (2021) propose a deep learning method for building a prediction model of vehicle emissions using parameters of both PEMS and OBD. In addition, PEMS data is often regarded as the benchmark technique. Zhang et al. (2023) focuses specifically on HDDVs, building a fuel-consumption based window method for $NO_x$ emission calculation by OBM data and validate the method within PEMS data. Similarly, numerous RSD studies have employed PEMS data for emission measurement validation (Ropkins et al., 2017, Bernard et al., 2022, Qiu and Borken-Kleefeld, 2022). Nonetheless, these studies encounter two primary challenges: (1) due to the complexity and high cost of PEMS tests, the sample size is often very limited; (2) the inaccessibility of OBM data presents a significant obstacle.

While OBM data may not track emissions from individual vehicles, it offers rich emission insights across multiple trips for specific vehicle types (Wang et al., 2022). By filtering data that satisfies predefined criteria and treating it as representative of a vehicle category, OBM data can elucidate dynamic emission profiles under varied driving conditions (Yang et al., 2016). This capability is crucial for establishing benchmarks to differentiate between standard and high-emitting vehicles. Conversely, RSDs have the capacity to capture the emissions of thousands of vehicles daily, positioning them as an efficient tool for initial vehicle screening (Ropkins et al., 2017). By leveraging the extensive coverage of remote sensing detection systems and the ongoing data sampling and analysis capabilities of OBM systems, a cost-effective emission surveillance framework could be built.

China leads globally as the first country to establish a national standard for remote sensing technologies, resulting in the widespread deployment of RSDs across major cities. Threshold of $1,500 \times 10^{-6}$ (ppm) for NO emission concentration is applied when screening out high-emitters (MEE, 2017). However, the concentration of emissions measured is dependent on the height of the tailpipe therefore simply using the NO concentration to identify candidate high-emitters is not very precise. In addition, the instantaneous emission of a vehicle varies across different driving conditions, a fixed threshold for diesel vehicles of different gross weight and different emission standard under all operating conditions has been debated. Further research is needed to develop a more robust method to effectively distinguish high $NO_x$ emitters from normally behaving vehicles. On the other hand, OBM have been integrated into the compliance framework since China VI (MEE and SAMR, 2018) for heavy-duty vehicles to identify faults and facilitate repairs and maintenance. In many cases now, however, OBM devices are only an alternative for pre-China VI HDDVs and that leads to frequent unavailability of OBM data. Combining OBM and RSD data to characterize emission performance of HDDVs and select high-emitters emerges as a crucial technological advancement.

This work proposes a novel method to screen out high-$NO_x$-emitters from HDDVs and estimates the corresponding emission reduction potential. Second-by-second profiles of $NO_x$ concentrations and driving conditions from the OBM data are collected to study the $NO_x$ emission performance at various vehicle operating conditions, and the relationship between instantaneous emissions and average emission levels is characterized. Then, the instantaneous

$NO_x$ emission recorded by the remote sensing device is linked to a corresponding emission distribution based on the driving condition. The possibility of a vehicle belonging to high-emitting vehicle subset is evaluated and the suspected high-emitters are screened out. At last, the results are generalized to estimate the $NO_x$ emission reduction potential that could result from identifying and repairing all vehicles with faulty, deteriorated or tampered emission aftertreatment systems.

## 2. Materials and methods

### 2.1. Data Acquisition and Processing

HDDVs for transporting construction waste (a.k.a. slag trucks) in Chengdu serve as an example in this paper. Slag trucks are a prevalent type of HDDVs in major urban areas in China, playing a crucial role in urban production activities and economic development. However, the concentration of construction sites within cities drives a high demand for these vehicles, contributing significantly to urban air pollution. In 2022, Chengdu had around 6,400 active daily slag trucks, with China V accounting for about 60-65% and China VI making up 35-40%. Many of these China V trucks face challenges related to aging aftertreatment systems which consequently results in a significant contribution to $NO_x$ emissions. Effectively identifying potential high-emitting vehicles among China V slag trucks could lead to a substantial reduction in air pollution.

Figure 1 describes the average seasonal road activity of China V slag trucks during the second to the fourth quarter of year 2022, with lighter shades highlighting areas of greater vehicle density. To analyze the spatial activity patterns of slag trucks, the investigated study area was segmented into 200m x 200m grids, with vehicles counted once per grid per season to map out their distribution. Figure 1 indicates the spatial distribution of China V vehicles has negligible variation over the three quarters in 2022. Due to constraints on computational power, the analysis of potential emission reductions in section 3.3 focused on the road activity of a typical weekday in May 2022, instead of encompassing an entire season or year. Figure 1's location markers illustrate the deployment of 13 remote sensing points across Chengdu. This extensive network of remote sensing devices underscores their capability for large-scale vehicle screening through RSD data. However, a limitation exists as a snapshot of emissions data captured by remote sensing can fluctuate with different driving conditions. This study uses OBM data that satisfies the data quality criteria for detailed characterization of $NO_x$ emissions from slag trucks. Leveraging this OBM data enhances the reliability and precision of identifying high-emission vehicles through remote sensing, offering a more effective approach to vehicle screening.

5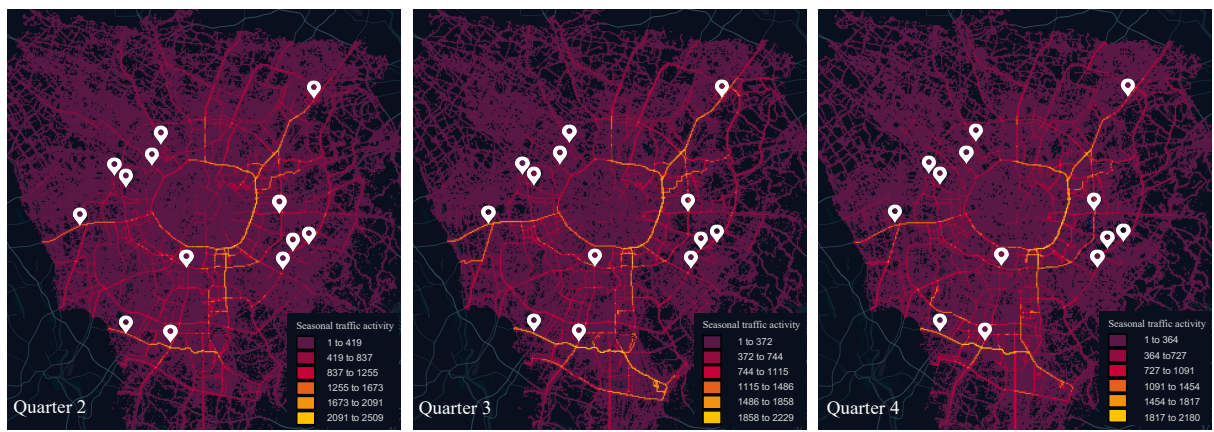

**Figure 1 seasonal activity data of China V slag trucks in Chengdu during Q2-Q4. Pins on the map indicate locations of remote sensing devices**

*2.1.1. OBM Data*

OBM is an emission control equipment that provides emissions-related diagnostics for the purpose of performance monitoring and analyzing repair needs. On-board diagnostic (OBD) systems as the predecessor of OBM were first installed on vehicles to monitor the status of engine and aftertreatment systems. And now in combination with the $NO_x$ sensors[1] and global positioning systems (GPS), OBD has evolved to OBM systems, becoming part of the future emissions compliance framework. Real-time engine information (engine speed & friction torque, engine fuel rate, etc.), $NO_x$ concentration, emission control system information (SCR temperature, DPF differential pressure, etc.), fuel tank level, velocity, longitude and latitude are reported remotely to the central management platform. HDDVs are equipped with OBM systems since China IV, which makes it a powerful and cost-effective method for emission monitoring of in-service vehicles.

The OBM data is used to determine the remote sensing cut points for high-emitters and calculate fuel consumption of slag trucks. This paper proposes to use the ratio of $NO_x/CO_2$ to evaluate the possibility of a vehicle belonging to high-emitting HDDV subset for the following reasons:

- Jiang et al. (2024) has proved that the ratio of $NO/CO_2$ has better correlation with real-world emission data than the concentration of NO (ppm); and
- The OBM measures the $NO_x$ emission concentration (ppm), and the RSD measures the NO emission concentration (ppm). To ensure the consistency between OBM emission data and RSD emission data, the ratio of $NO_x/CO_2$ is considered a more appropriate metric to screen out high-emitting vehicles. A fixed fraction of primary $NO_2$ is used to estimate the concentration of NO$_x$ (NO + NO$_2$) in the RSD data.

---

[1] $NO_x$ sensor is monitored before and after the Selective Catalytic Reduction (SCR) system to measure the $NO_x$ being emitted from the diesel engine, and whether the SCR is working efficiently to remove $NO_x$.

To derive the ratio of NO$_x$/CO$_2$ from the OBM data, the instantaneous emission rate of NO$_x$ ($g/s$) and CO$_2$ ($g/s$) are calculated by the following equation: The fuel consumption ($L/km$) is calculated by equation (4) (Zhao, 2022):

$$ER_{NO_x} = \frac{\mu \times NO_{x_{out}} \times q_t}{3600} \tag{1}$$

$$q_t = q_{MAF} + q_{FR} \times \rho \tag{2}$$

$$ER_{CO_2} = \frac{q_{FR} \times \beta}{3600} \tag{3}$$

where $ER_{NO_x}$ is the instantaneous NO$_x$ emission rate of the vehicle at time $t$ ($g/s$); $\mu$ represents the ratio of exhaust component density to exhaust density (the proportion of NO$_x$ in the total tailpipe emissions), which is set to 0.001587 for diesel fuels (Wang et al., 2022); $NO_{x_{out}}$ is the concentration of NO$_x$ after SCR (ppm); $q_t$ is the mass flow rate of NO$_x$ emissions at second t ($kg/h$); $q_{MAF}$ is the engine intake flow rate at time $t$ ($kg/h$); $q_{FR}$ is the engine fuel rate at second ($L/h$); $\rho$ is the fuel density of the engine ($kg/L$), which is set to 0.85 for diesel fuels (Wang et al., 2022); $ER_{CO_2}$ is the instantaneous CO$_2$ emission rate of the vehicle at second t (g/s); $\beta$ is the grams of CO$_2$ produced when burning one litre of fuel ($g/L$), which is set to 2684 for diesel fuels (ICCT, 2022).

The regulatory agency mandates that the data be reported at intervals of no more than 10 seconds (MEE and SAMR, 2018) since China VI. However, for China V slag trucks, the OBM data is frequently missing, which can impact the effective evaluation of NO$_x$ emission performance. To guarantee data quality, an OBM data sample selection and processing method has been developed and is stated as follows:

(1) The service time for each trip should last a minimum of 30 minutes, and records with time intervals longer than 12 seconds should constitute no more than 30% of the entire trip.

(2) Invalid data points should account for less than 30% in one trip. Invalid data are defined as data necessary for calculating the instantaneous emission rate of NO$_x$ ($g/s$) fall outside the value range set by MEE and SAMR (2018). Additionally, if any required data show no variation for 10 consecutive records, they are considered invalid.

(3) The OBM data is used to calculate the acceleration of a vehicle at second t in section 2.2. The sampling frequency of vehicle speed in the OBM data is 10 seconds and accurate acceleration cannot be derived based on it (Piccoli et al., 2015). The imputation methods based on random forests has been validated and is used for acceleration prediction in this paper (Hong and Lynn, 2020).

At last, 76 China V slag trucks (1,713 trips in total) are accessed and analyzed after the OBM data selection process in this paper, covering a period of ten months from February 2022 to December 2022.

7*2.1.2. RSD Data*

A remote sensing system is positioned at the roadside (Bishop and Stedman, 1996, Huang et al., 2018) or above the roadway (Ropkins et al., 2017, Ghaffarpasand et al., 2020). It passes infrared (IR) and ultraviolet (UV) light beams through the exhaust plume of the passing vehicles, and by measuring the absorption of the light wavelengths, the concentration of pollutants in the exhaust plume is derived. Speed and acceleration when the vehicle passing by the measurement location, vehicle information (vehicle type, fuel type, emission standard, etc.) are also recorded together with the snapshot of vehicle emissions. RSDs offer flexible spatial coverage and can efficiently sample vehicles as they pass by, at a relatively low cost.

When an RSD is used for on-road vehicle emissions testing, Dilution of exhaust gases with ambient air can occur and this can affect the absolute concentrations of pollutants. To mitigate the impact of dilution, this paper therefore uses the ratio of $NO_x/CO_2$ as the metric to screen out potential high-emitting HDDVs, as both NO and CO2 are influenced by it to some extent. The ratio of $NO_x/CO_2$ is computed by the following equation:

$$\frac{NO_x}{CO_2} = \frac{NO/(1-f_{NO_2})}{CO_2} \qquad (4)$$

where $f_{NO_2}$ is the fraction of $NO_2$ in $NO_x$, which is set to 40.0% for China V HDDV (Lau et al., 2015) in this study.

A collection of 3,400 remote sensing measurements from China V diesel slag trucks are selected and analysed. These records are from regular remote sensing screening program from January 2023 to July 2023 in Chengdu across 2 sampling sites. The detailed site information is stated in Table 1. RSD-1 is a cross-road device while RSD-2 is a top-down device; beyond this, both sites share comparable measurement conditions.

**Table 1 Remote Sensing Monitoring Site Information**

| Site ID | Count | Road grade (%) | Median. Speed ($km/h$) | Median. Accel ($m/s^2$) | Median. Temp (°$C$) |
|---|---|---|---|---|---|
| RSD-1 (cross-road) | 404 | 0 | 49.0 | 0.3 | 23.4 |
| RSD-2 (top-down) | 2,996 | 0 | 52.0 | 0.3 | 24.9 |

**2.2. Method**

Vehicle Specific Power (VSP) value is a metric informing the power demands on the engine during driving (Jiménez-Palacios, 1999). The VSP value is often considered to be associated with $NO_x$ emissions (Carslaw et al., 2013, Yang et al., 2016). It is used to identify whether vehicles are at high load (where high emissions are expected) or at very low load (where fuel injection is disabled and plume sizes insufficient for valid remote sensing measurements). To ensure the comparability between the $NO_x$ emissions from the RSD measurements and from the OBM data, the VSP value is used as a metric to label NO$_x$ emissions in this paper. The VSP value is calculated using the second-by-second speed and acceleration values in a driving schedule, along with information about the type of vehicle being operated:

$$VSP_t = a_t \times v_t + g \times v_t \times \sin \theta_t + \frac{A}{m} \times v_t + \frac{B}{m} \times v_t^2 + \frac{C}{m} \times v_t^3 \tag{5}$$

where $VSP_t$ is the vehicle specific power at second t ($kW/ton$); $a_t$ is the vehicle acceleration at second t ($m/s^2$); $v_t$ is vehicle speed at second t ($m/s$); $g$ is the gravitational acceleration, which is approximately $9.8\ m/s^2$; $\theta_t$ is the road gradient at second t (%), and $\theta_t$ is considered as 0 in this paper; m is the vehicle mass including loading ($ton$); A is the rolling resistance coefficient ($kWs/m$); B is the rotational resistance coefficient ($kWs^2/m^2$); $C$ is the aerodynamic drag coefficient ($kWs^3/m^3$);. The road-load coefficients (i.e., $A/M$, $B/M$, and $C/M$) of HDDVs used in this study are 0.0875, 0, and 0.000331 (Wu et al., 2012).

This paper identifies high-emitters and computes emission factors in separate operating model bins, which could offer the opportunity to characterize the emission performance of HDDVs in a higher resolution. Based on the GPS profiles of HDDVs in Chengdu and test cycle of slag trucks (CHTC-D) from the latest China automotive cycle standard (SAMR and SAC, 2019), the operating modes are divided by the instantaneous vehicle speed and VSP value (see Table 2). A total of 22 modes are developed including braking mode (Bin 0), idle mode (Bin 1), low-speed modes (Bin 11-16) ($1.6\ km/h \leq v < 30\ km/h$), medium-speed modes (Bin 21-28) ($30\ km/h \leq v < 60\ km/h$) and high-speed modes (Bin 33-38) ($v \geq 60\ km/h$).

Table 2 Operating mode bins by vehicle speed and VSP values

| VSP (kW/ton) | | Vehicle speed (km/h) | | | |
|---|---|---|---|---|---|
| | | v < 1.6 | 1.6 ≤ v < 30 | 30 ≤ v < 60 | v ≥ 60 |
| VSP < −4 | | | Bin 11 | Bin 21 | |
| −4 ≤ VSP < −2 | | | Bin 12 | Bin 22 | Bin 33 |
| −2 ≤ VSP < 0 | | | Bin 13 | Bin 23 | |
| 0 ≤ VSP < 2 | Bin 0 Braking | Bin 1 Idle | Bin 14 | Bin 24 | Bin 34 |
| 2 ≤ VSP < 4 | | | Bin 15 | Bin 25 | Bin 35 |
| 4 ≤ VSP < 6 | | | | Bin 26 | Bin 36 |
| 6 ≤ VSP < 8 | | | Bin 16 | Bin 27 | Bin 37 |
| VSP ≥ 8 | | | | Bin 28 | Bin 38 |

Note: The definition of the braking condition: the instantaneous acceleration at second t is less than $-0.89\ m/s^2$.

To evaluate the emission performance of slag trucks, the distance-specific NO$_x$ emission factors is derived by combining the fuel-specific NO$_x$ emission factor calculated based on the RSD data and the fuel consumption calculated based on the OBM data.

$$EF_{(\text{Distance-specific})NO_x} = EF_{(\text{Fuel-specific})NO_x} * FC * \rho \tag{6}$$

$$EF_{(\text{Fuel-specific})NO_x} = \frac{30 * Q_3 * 860}{(1 + Q_1 + 6Q_2) * 12} \tag{7}$$

$$FC = \frac{\sum_{j=Bin0}^{Bin38}(q_{FR} \times P_j)}{\bar{v}_j} \tag{8}$$

where $EF_{(\text{Distance-specific})NO_x}$ is the distance-specific NO$_x$ emission factor ($g/km$); $EF_{(\text{Fuel-specific})NO_x}$ is the fuel-specific NO$_x$ emission factor ($g/kg$); $Q_1$ is the ratio of CO to CO$_2$; $Q_2$ is the ratio of HC to CO$_2$; $Q_3$ is the ratio of NO$_x$ to CO$_2$; $EF_{(\text{Fuel-specific})NO_x}$ is the fuel-specific NO$_x$ emission factor ($g/kg$); $FC$ is the average fuel consumption of the vehicle (L/km); $q_{FR}$ is the engine fuel rate in Bin j ($L/h$); $P_j$ is the percentage of operating time in Bin j; $\bar{v}_j$ is the average speed in Bin j ($km/h$).

## 3. Results and discussions

### 3.1. Characterization of NOx Emissions Based on the OBM Data

Characterizing the real-world driving conditions is essential when analyzing the vehicle emissions. The amount of time spent in each of the 22 operating mode bins based on the OBM data indicates the typical driving pattern of slag trucks in Chengdu (shown in Figure 2(a)). The operating mode bins follow the normal distribution in all three speed ranges (Bin 11-16 for low-speed range, Bin 21-28 for medium-speed range, Bin 33-38 for high-speed range), and the slag trucks are mostly driven at low-speed range and medium speed range. Figure 2(b) is a demonstration of the driving characteristics of slag trucks when passing by a remote sensing device, where the operating mode is more centralized in Bin 28. The preferred placement of RSDs for vehicular emissions monitoring is in areas with higher vehicle speeds, as this promotes more consistent engine loads. Consequently, the resulting emissions measurements are more uniform and less affected by driving pattern variability.

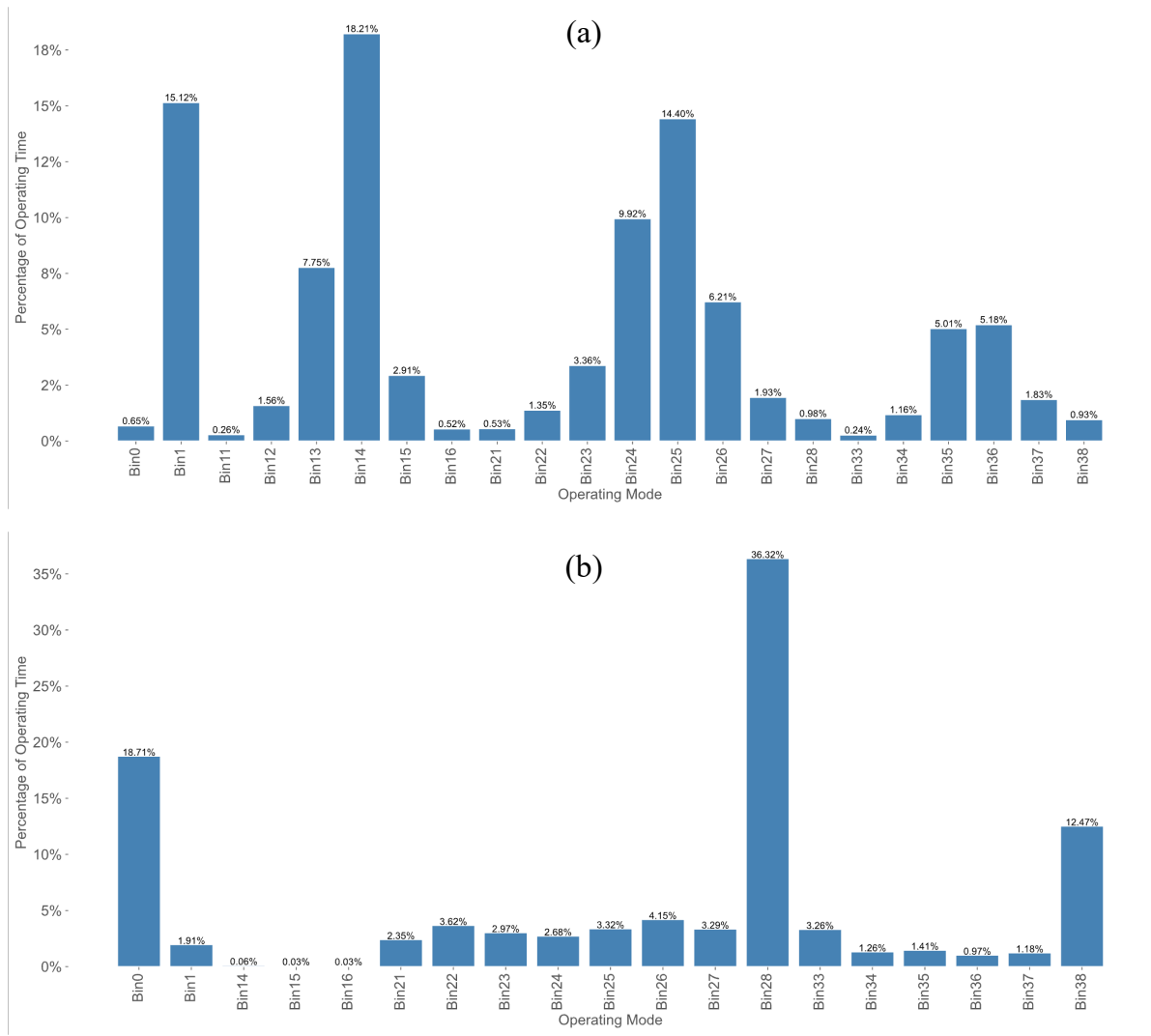

**Figure 2 Distribution of Operating Mode Bins based on (a) OBM Data and (b) RSD Data**

The ratio of $NO_x/CO_2$ for determining the high-emitter threshold is calculated based on the OBM data. The average value in each operating bin is used to represent the typical emission performance of normally behaving vehicles in one specific driving condition, and the high-emitter thresholds are set to two times the corresponding average value (Qiu and Borken-Kleefeld, 2022). Figure 3 shows the average emission performance in each operating bin. The median-speed range and high-speed range has a clear trend that the NOx emissions is getting lower as the VSP value increases. The prevailing theory suggests that elevated, consistent engine load may generate sufficient heat to keep the NOx emission control systems, such as Exhaust Gas Recirculation (EGR) and Selective Catalytic Reduction (SCR) used in diesel engines, within their optimal temperature range for high-efficiency operation (Yang et al., 2022). In comparison, in the low-speed range, no evident trends of the typical emission rates are observed in the collected OBM data. This may be due to that emission performance in the low-speed range might be affected by many factors such as the cold start (Chu Van et al., 2019).

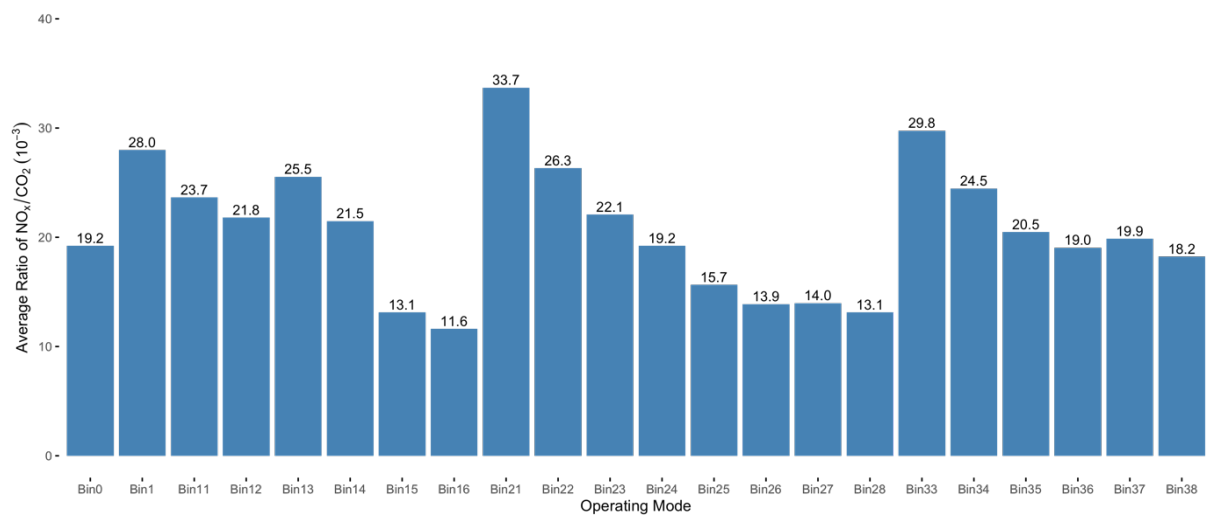

**Figure 3 Boxplot of Ratio of $NO_x/CO_2$ in Each Operating Mode Based on OBM Data**

### 3.2. Identification of Candidate High-emitting Vehicles

The threshold determined by the OBM data is utilized to identify potential high-emitters in the RSD data. Vehicles passing by remote sensing devices are categorized into different operating modes based on the instantaneous speed and acceleration. The ratio of $NO_x/CO_2$ computed based on the RSD data is compared with the corresponding high-emitter threshold to screen out candidate high-emitting vehicles. It's crucial to emphasize that, in accordance with the national standard, a vehicle's NO concentration must surpass the threshold at least twice within a 6-month period to qualify as a high-emitter. This identical criterion is also employed in the novel OBM+RSD method. However, due to the limited volume of available RSD data, a vehicle is regarded as a potential high-emitter if it exceeds the threshold in any of the three speed ranges at least twice, as opposed to the ideal 22 operating bins. Out of the 3,400 RSD measurements, 956 unique vehicles have been recorded. According to the national standards, 18 vehicles surpassed the emission limit of 1,500 ppm at least twice, and that represents 1.9% of the China V slag trucks evaluated are high-emitters. When utilizing the combined OBM and RSD approach, number of vehicles and high-emitter percentage in each of the speed range are stated in Table 3, where around 7% of the entire fleet are examined as potential high emitters. Given that the low-speed range includes only four vehicles, its high-emitter percentage is reflected as the average high-emitter percentage of the entire dataset. The result indicates that there is a potential leniency in the national threshold for screening out high-emitting in-use diesel vehicles by remote sensing method.

To achieve a more detailed characterization of the emission performance for both normally-behaving vehicles and high-emitting vehicles, the emission factors for slag trucks have been estimated across three speed ranges. Distance-specific $NO_x$ emission factors (g/km) are calculated by combining the fuel-specific $NO_x$ emission factor (g/kg) from the RSD data, the average fuel consumption (L/100km) from the OBM data, and the fuel density of diesel (0.85 kg/L). For China V slag trucks, the average fuel consumption is calculated at 51.5 L/100km, as per equation (10). Table 3 details the corresponding distance-specific NO$_x$ emission factors of both high-emitters and normally-behaving vehicles in different speed ranges. The NO$_x$ emission factor of normally-behaving vehicles in the medium-speed range is the same as in the high-speed range, while the NO$_x$ emission factor of high-emitters in the medium-speed range is lower than in the high-speed range. The findings indicate that vehicles with operating aftertreatment systems have a stable emission performance in various driving conditions, while vehicles with impaired, degraded, or tampered with emission aftertreatment systems exhibited poorer emission performance when the engine load is high (in high-speed range).

Table 3 High-emitter percentage and the distance-specific NO$_x$ emission factors (g/km)

| Speed range | Total vehicles | High-emitter percentage | Emission factor of NBVs | Emission factor of HEs |
|---|---|---|---|---|
| **By the OBM+RSD method** | | | | |
| Low-speed range (1.6-30 km/h) | 4 | -- | -- | -- |
| Medium-speed range (30-60 km/h) | 712 | 6.9% | 4.8 | 14.2 |
| High-speed range (≥ 60 km/h) | 295 | 2.7% | 5.3 | 16.2 |
| **By the national threshold** | | | | |
| Total | 956 | 1.9% | 5.7 | 16.4 |

### 3.3. Emission reduction potential of multi-data vehicle monitoring

Combining the knowledge gained in previous sections about the NO$_x$ emission performance of both normally behaving slag trucks and candidate high-emitters in each operating bin, with available vehicle traffic statistics for the China V fleet, the results are generalized to estimate the NO$_x$ emission reduction potential achievable by identifying and repairing all high-emitting vehicles to match the performance of those operating normally. Figure 4(a) and Figure 4(b) respectively illustrate the total emission reduction potential during daytime (8:00-20:00) and night-time (20:00-8:00) periods. With construction-related transport more common during daytime (Figure 4(a)), estimated total emission reductions from 8:00 to 20:00 amount to 153kg. At night-time, the removal of traffic restrictions increases vehicle activity in city centres (Figure 4(b)), with total emissions reductions during night-time off-peak hours (20:00 to 8:00 the next day) reaching 97kg. Overall, daily relative NO$_x$ emission savings in comparison to current levels are calculated at 18.8%, or 250kg. The emission reduction achieved by identifying high-emitters therefore significantly reduces the population's exposure to high NO$_x$ concentrations. Figure 4(c) illustrates the potential per-vehicle emission reduction possible through the combined use of OBM+RSD approach, emphasizing a notable decline in $NO_x$ emissions on major roads with speeds above $30 km/h$.

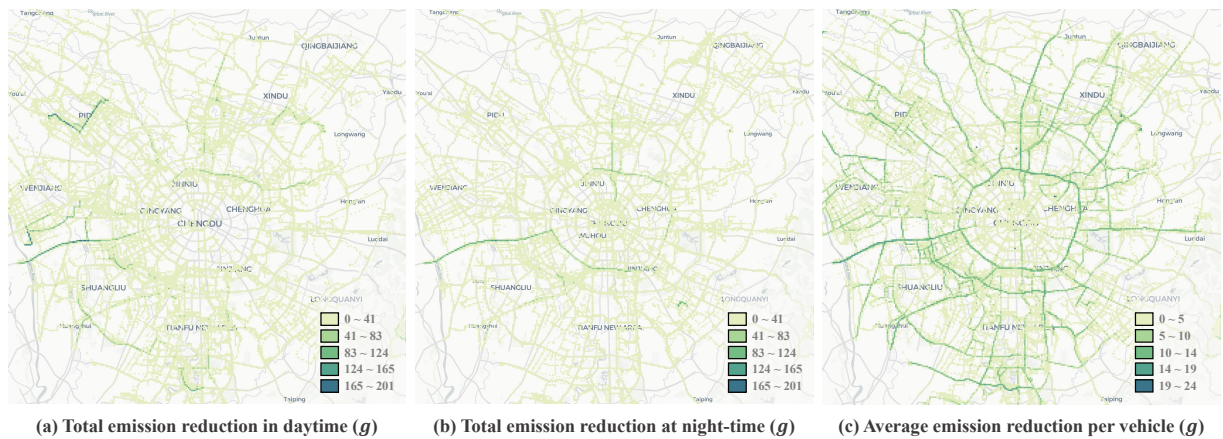

(a) Total emission reduction in daytime ($g$)   (b) Total emission reduction at night-time ($g$)   (c) Average emission reduction per vehicle ($g$)

**Figure 4 (a) daily total emission reduction during daytime peak hours (8 am to 8 pm); (b) daily total emission reduction during night-time off-peak hours (8 pm to 8 am); (c) daily average emission reduction per vehicle**

The emission map delineates key areas for targeted emission control efforts during both daytime and night-time, underscoring the necessity of adapting strategies to the temporal dynamics of urban traffic and pollution patterns. While the current study focuses on China V slag trucks, extending this approach to other vehicle categories and emission standards (e.g., China VI) could further enhance its applicability and impact. Additionally, exploring the scalability of the OBM+RSD approach to other urban centres and varying traffic and environmental conditions could provide deeper insights into its broader implementation potential.

## 4. Conclusion

HDDVs with a gross weight ranging from 12 tonnes to 25 tonnes are utilized across various services, including sanitation, construction, delivery, etc. These vehicles are granted access to central urban areas, indicating their potential significance in contributing to the urban air pollution. This paper combines the OBM data with the RSD data to identify candidate high-emitters and derive typical $NO_x$ emission factors of China V slag trucks. The OBM data is used to demonstrate the typical emission characteristics of slag trucks and determine the high-emitter threshold under real-world driving profiles. By using the metric of $NO_x/CO_2$ rather than the $NO_x$ concentration, a more robust insight of the real emission performance is provided. RSD measurements are used to screen out candidate high-emitters in the China V fleet based on corresponding thresholds decided by the OBM data. The emission reduction potential is estimated by combining the road activity, emission factors, and high-emitter percentage in different speed ranges. The key findings of the paper are:

- There is a potential leniency in the national threshold for screening out high-emitting in-use diesel vehicles by remote sensing method. It is suggested that thresholds are

- established based on the real-world emissions and driving conditions rather than a fixed point.
- High-emitters exhibited poorer emission performance in the high-speed range, however the high-emitter percentage of these vehicles are lower in the high-speed range.
- There is no difference in the high-emitter percentage of the two remote sensing sites used in this study. With the deployment of more RSDs, this information could help the local authorities to enhance monitoring of in-use diesel vehicles in certain areas.
- A fleet-average reduction of 18.8% is found if all high-emitting China V slag trucks in Chengdu have been repaired and have the same emission performance as the normally-behaving vehicles. Major roads in Chengdu have seen the largest substantial reduction in $NO_x$ emissions. The same method can be extended to China VI slag trucks or other HDDVs for higher emission reduction.

This research introduces a novel method to screen candidate high-emitting diesel vehicles and contributes to the knowledge of the real-world emission performance of HDDVs. Although it provides a promising tool for emission monitoring, it still has some drawbacks. The estimation of emission reduction potential relies on the assumption that all China V slag trucks have been detected at least once by the RSDs. However, at the time of remote sensing data collection, only two remote sensing sites were operational. To optimize the effectiveness of using the RSDs for screening high-emitting vehicles, the deployment density of RSDs needs enhancement, and the regular operation of the RSDs must be ensured. Additionally, further research is deemed necessary to develop a more robust method for characterizing fleet behavior and effectively distinguishing high-emitters in operating bins that cannot be characterized by the Gumbel distribution. The estimation of emission reduction potential relies on distance-specific emission factors within three speed ranges, rather than 22 detailed operating bins, due to a lack of data. As more OBM and RSD data are collected, it would be possible to map the $NO_x$ emissions of China V slag trucks in Chengdu in higher resolution.

Emission monitoring of in-service vehicles has long been the focus of urban air pollution control. Local governments like Shanghai have recently published pilot programs to further enhance the in-use compliance for HDDVs. Vehicles equipped with OBM systems that are both networked and consistently attains compliance with corresponding emission limits, may be eligible for exemption from the periodic emission inspection process. However, researchers have also discovered that even when the OBM system is operating as expected, there is still a possibility that the emissions from the tailpipe are above the limit (Jiang et al., 2021). In addition, OBMs as part of the emission compliance framework have been implemented on HDDVs since China IV, however due to data communication problems, the OBM data quality may not be reliable for the purpose of emission monitoring. In that case, the OBM alone is not enough for effective monitoring. Technologies are emerging (e.g., annual I/M Program, RSD, OBM, on-road random inspection) for the purpose of monitor the real-world emission condition of diesel vehicles and it is challenging to combine the heterogeneous multi-source data to establish a comprehensive and continuous emission monitoring system. OBM data has long temporal coverage and is able to characterize the real-world emission performance, while RSD data has flexible spatial coverage and can sample vehicles whenever the vehicle is passing by, especially those China V HDDVs without adequate OBM data. Leveraging the complementary features of these two monitoring methods is an ideal and cost-effective approach to effectively identify candidate high-emitters. This paper provides a solution and practice to combine a snapshot of emission records and time-specific emission rates, improves the validity required of using a single RSD record to screen out high-emitters, and assess the emission reduction of HDDVs in higher resolution.

## Acknowledgments

This work was supported by the Fundamental Research Funds for the Central Universities (no. 2682023CX044) and National Natural Science Foundation of China (no. 72071163).